# Predicting 3D RNA Folding Patterns via Quadratic Binary Optimization

Mark Lewis  mlewis14@missouriwestern.edu
Craig School of Business, Missouri Western State University, Saint Joseph, MO  64507

Amit Verma averma@missouriwestern.edu
Craig School of Business, Missouri Western State University, Saint Joseph, MO  64507

Rick Hennig  rick@meta-analytics.com
Meta-Analytics, Inc. Boulder, CO 80302

**Abstract**.  The structure of an RNA molecule plays a significant role in its biological function. Predicting structure given a one dimensional sequence of RNA nucleotide bases is a difficult and important problem.  Many computer programs (known as *in silico*) are available for predicting 2-dimensional (secondary) structures however 3-dimensional (tertiary) structure prediction is much more difficult mainly due to the far greater number of feasible solutions and fewer experimental data on the thermodynamic energies of 3D structures.  It is also challenging to verify the most likely three dimensional structure even with the availability of sophisticated x-ray crystallography and nuclear magnetic resonance imaging technologies.  In this paper we develop three dimensional RNA folding predictions by adding penalty and reward parameters to a previous two dimensional approach based on Quadratic Unconstrained Binary Optimization (QUBO) models.   These parameters provide flexibility in the amount of three dimensional folding allowed.  We address the problem of multiple near-optimal structures via a new weighted similarity structure measure and illustrate folding pathways via progressively improving local optimal solutions.   The problems are solved via a new commercial QUBO solver AlphaQUBO (Meta-Analytics, 2020) that solves problems having hundreds of thousands of binary variables.

**Introduction**.  The ribonucleic acid (RNA) folding problem is the modeling and prediction of the ending shape (structure) with a linear input sequence of nucleotides {A, C, G, U} folding back on itself to form secondary and tertiary structures (see Figure 1).  The beginning unpaired sequence of bases is a one dimensional strand of RNA, and a strand that then folds back onto itself via connections with complementary base pairs forms a secondary, two dimensional structure.   During folding an RNA may also acquire three dimensions, forming what are called pseudoknots.  The prediction of a 3D structure from a linear sequence of bases is a very complex undertaking and is the subject of this paper.

*In silico* is a term referring to the use of computer algorithms and simulations to investigate biological interactions vis-à-vis experiments performed inside cells in a laboratory (*in vivo*) or in a natural setting (*in situ*).  Predicting *in silico* the feasible and most likely structures of a sequenced RNA is useful in reducing the time and costs associated with physically testing alterations in base sequences as part of drug discovery and repurposing.  For example, *in silico* sequence analysis and binding site flexibility have been used to support drug repurposing of FDA approved anti-viral and anti-malarial drugs for use against Covid-19 (Lee, et al., 2020).   Predicting RNA folding pathways and stable structures allows the discovery of nucleotide binding sites to which interacting molecules, such as in the form of therapeutic drugs, can be attached.

**Literature**. The accurate modeling of RNA's thermodynamic and chemical properties is critical to accurate prediction results. This paper's model utilizes thermodynamic base pair nearest neighbor bonding energies based on (Turner & Mathews, 2009) along with primary constraints such as a minimum distance between base pairs, a base at position *i* can only pair with at most one other base and restriction to the canonical Watson-Crick base pairings AU, GC and GU. There are other energy models for base pairing, as well as energy data associated with common structural motifs, and not all RNA prediction programs use the same underlying energy information. Other methods for structure prediction are Comparative Sequence Analysis based on evolutionary evidence that similar RNA from similar organisms reliably form the same structures (Cannone, et al., 2002). Base-pairing probability matrices are another important tool that can supplement other structure prediction approaches (Hamada, 2012).

Specific structural characteristics, or motifs, include hairpins, loops and bulges are used by software to determine the minimum free energy (MFE) of a structure, where a minimal free energy implies it is in a stable state. Figure 1 illustrates the folded RNA structural characteristics commonly encountered in 2D folding as well as the kissing hairpin (lower right), a common motif for pseudoknots, generally defined as an overlapping in the linear sequence of stems.

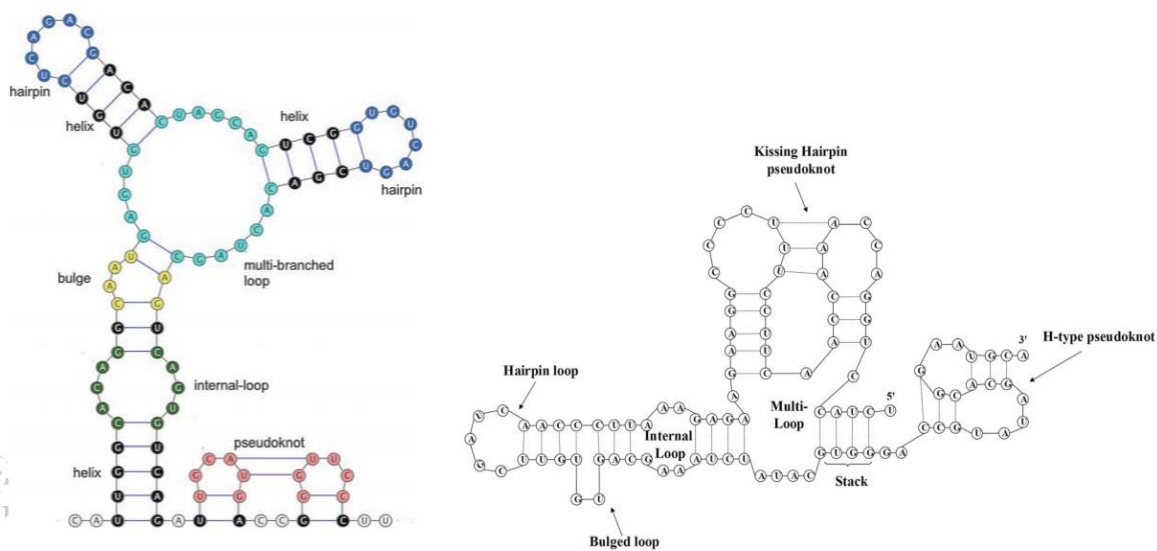

Figure 1. Common 2 and 3 dimensional RNA structures from (Mamuye, et al., 2016) left, and (Kai, et al., 2019) right.

As a point of distinction, our software approach is conceptually the mathematical dual of minimizing free energy because it maximizes the formation of stems via sequential base pairings and the hairpins, loops, bulges and other features are created during this process. A similar approach focusing on the bond energy in stems is presented in (Gupta, et al., 2012). The same article raises an important point, that prediction of (and comparison to) a single best structure falls short of capturing the dynamics of the RNA folding landscape and that many existing approaches provide a set of structures to the decision maker.

When evaluating a prediction, laboratory techniques such as Nuclear Magnetic Resonance (NMR) and X-ray Crystallography provide important benchmarks for RNA structures. However, they are expensive, sensitive, take time and expertise and affect RNA viability because they require it to be modified in preparation for measurement, e.g., adding a fluorine solution for NMR, or molecular crystallization for X-ray analysis. While NMR has been restricted to less than 35 bases, recent progress on RNA with 61 bases shows promise for larger molecules (Becette, et al., 2020). Although X-ray crystallography can identify the structure of RNA with over 1000 bases (if they can be crystallized), NMR has an advantage in that it can reveal the RNA as it changes shape, known as the folding pathway. A folding pathway progressing towards a stable state is important as binding sites more amenable to therapeutic drugs may appear during the folding process. While *in silico* RNA prediction software evaluation compares the ability to match a benchmark structure, it is important to understand that the RNA folding process is dynamic and the stable state associated with the lowest MFE is not necessarily stable. Figure 2 contains two benchmarks available for the same molecule and illustrates the structural differences that can occur between a 2D and 3D benchmark.

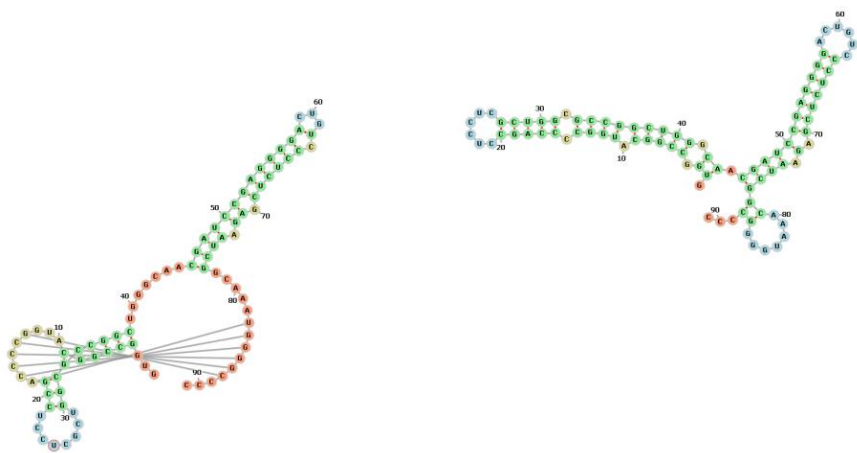

Figure 2. Two benchmarks for molecule RFA_00632 showing the 3D folded structure from (RNA STRAND Database, 2008) with pseudoknots on the left and the 2D structure on the right from (Vienna RNA Web Service, 2020).

Minimizing thermodynamic free energy (MFE) is an essential component of 2D folding prediction, but it is not particularly good at predicting 3D pseudoknots, mainly because the free energy measurements used in the calculations are based on 2D structures (not 3D) (Gardner & Giegerich, 2004). The minimum free energy data for 3D structures with pseudoknots are not widely available, so that in this paper, comparisons between our structure predictions and X-ray or NMR 3D benchmarks will be using similarity measures based on Accuracy, Precision, Sensitivity and Matthews Correlation Coefficient instead of MFE. In addition, a new measure designated Weighted Most Similar Structure will be introduced to resolve issues associated with selecting from a set of elite structures.

The gaps between *in vitro* and *in vivo* RNA folding investigations are presented in (Leamy, et al., 2016). They also describe how *in silico* analysis informs the other approaches by providing alternate hypotheses to investigate. The importance of RNA folding landscapes and pathways is discussed in (Kucharik, et al., 2016) which also states that pseudoknots play an important role in folding trajectories but are usually excluded because of the added combinatorial complexity, as well as difficulties inherent in deriving free energy metrics in their presence. Dynamic programming approaches based on minimum free energy are fast at 2D prediction, but they break down when pseudoknots are present. Dynamic programming and other approaches are discussed in (Bellaousov & Mathews, 2010) wherein they propose ProbKnot (Mathews Group, 2021) for predicting secondary structures in the presence of pseudoknots.

**3D Folding Model**. QUBO models have the general form

$$\text{Max: } \sum_i^n c_i x_i + \sum_i^n \sum_j^n c_{ij} x_i x_j \text{, or equivalently Max } x'Qx$$

where x is an array of binary variables and Q is a symmetric $n \times n$ matrix of coefficients $c_{ij}$. The definition and meaning of the binary variables $x_i$, the linear coefficients $c_i$ and interacting quadratic coefficients $c_{ij}$ are critical to accurate modeling. The diagonal of Q contains linear terms $c_i$ that quantify the effect of flipping a single variable from 0 to 1 and the off-diagonal quadratic coefficients $c_{ij}$ represent the magnitude of the interaction effect when both variables $x_i$ and $x_j$ are set to one. Without loss of generality, $x_i x_j = x_j x_i$ resulting in a symmetric Q matrix where we need only define pairs $x_i$ and $x_j$ for $i < j$.

QUBO, by definition, are unconstrained, however constraints can be implemented as penalties in the objective function where the magnitude of the penalty determines how strictly the constraint is enforced. A small penalty magnitude would allow a constraint to be violated if it produced benefit. Larger penalty magnitudes will not allow constraint violation. Minimum Free Energy is the common metric in 2D thermodynamic models, however MFE is not available for 3D prediction and in this research the objective function $xQx$ is optimized as a proxy for maximizing the number of paired bases.

In the paper (Lewis, et al., 2021) the binary variables $x_k$ are defined as representing feasible base pairings between *two* pairs, known as a stacked quartet. A single base pair is not stable (Kai, et al., 2019) thus the simplest binary variable is defined as a stacked quartet (two nested pairs). Let the variable $x_k$ represent a sequential *nested* pair (stacked quartet) of 4 bases at positions in the linear sequence $i^1, i^2, j^1, j^2$, hence $x_k^{i^1,j^1,i^2,j^2} \in \{0,1\}$ indicate the pairs $(i^1, j^1)$ and $(i^2, j^2)$ are both feasible pairs and nested, i.e. $i^1 + 1 = i^2$ and $j^2 + 1 = j^1$. The superscripts are removed for brevity in future references. Referring to Figure 2, the variable $x_s = ((1, 10), (2, 9))$ and the variable $x_t = ((2, 9), (3, 8))$ together represent a stem (helix) of three sequential nested base pairs. Selecting both these variables would be rewarded in the model via $c_{ij}$ because they lengthen a short stem, increasing bound energy. The linear coefficients in the Q matrix represent the binding energies (Turner & Mathews, 2009) of a single nested pair $x_k$ and are provided in Figure 3.

|   |    | Stacked pair 5' to 3' |     |     |     |     |     |
|---|----|-----|-----|-----|-----|-----|-----|
|   |    | AU  | CG  | GC  | UA  | GU  | UG  |
|   | AU | 0.9 | 2.2 | 2.1 | 1.1 | 0.6 | 1.4 |
|   | CG | 2.1 | 3.3 | 2.4 | 2.1 | 1.4 | 2.1 |
| 5'| GC | 2.4 | 3.4 | 3.3 | 2.2 | 1.5 | 2.5 |
| to  | UA | 1.3 | 2.4 | 2.1 | 0.9 | 1   | 1.3  |
| 3'  | GU | 1.3 | 2.5 | 2.1 | 1.4 | 0.5 | -1.3 |
|     | UG | 1   | 1.5 | 1.4 | 0.6 | -0.3| 0.5  |

Figure 3. Stacked Quartet free energies (kcal/mol) from (Turner & Mathews, 2009)

There are three options for the Q matrix quadratic interaction terms $c_{ij}$ that define the relationship between two stacked quartets $x_k$ and these are M⁺, M⁻ and M^P. M⁺ is a reward used to promote nesting and sequencing with other nested pairs in order to generate long stable stems. M⁻ is used to discourage crossing nested pairs (pseudoknots) and represents a soft constraint because by making |M⁻| < |M⁺| the violation of the soft constraint, in order to form a longer stem, may increase the objective function value. M^P is a large penalty that prevents assigning the same base to a different matching pair. Hence, |M⁻| < |M⁺| < | M^P|.

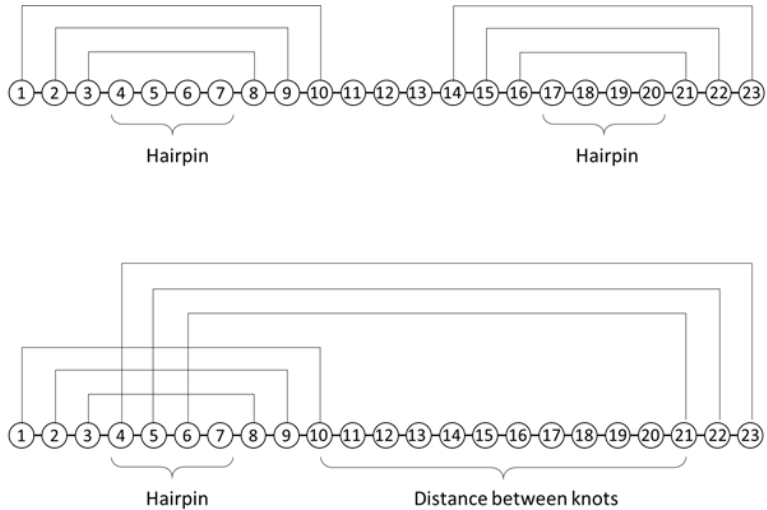

Figure 4. Non-crossing base pairs (top) has two hairpins with unpaired genes 4-7 and 17-20. Pseudoknot with bases (4,5,6) in hairpin crossing to pair with bases (23,22,21)

In summary, the Q matrix is composed of linear terms representing the binding energy of stacked quartets represented by variables $x_k$. These variables are discovered and enumerated by examining the input linear RNA sequence for feasible base pairs. A single base can be included in many base pairs and thus many $x_k$ variables and the penalty to prevent assigning the same base to two different variables with different base pairs, $x_s$ and $x_t$, is given by M^P hence $c_{st}$ = M^P. Pseudoknots are defined as crossing base pairs $(i, j)$ and $(i', j')$, where $(i < i' < j < j')$. Any $x_s$ and $x_t$ whose base pairs meet these criteria are penalized in the Q matrix with magnitude $c_{st}$ = M⁻. In Figure 4, if $x_s$= ((2,8), (3,9)) and $x_t$ = ((4, 23), (5, 22)) then they create a pseudoknot because (2, 8) and (4, 23) meet the crossing pair criteria of 2 < 4 < 8 < 23.

The fundamental building blocks of both DNA and RNA are nucleotides (nt), referred to as bases, composed of a sugar, a phosphate, and a nitrogenous base. The primary chemical difference between DNA and RNA is the type of sugar used, which is deoxyribose for DNA and ribose for RNA. Nucleotides are strung together to form DNA or RNA polynucleotides by chemical reactions that result in a backbone of alternating sugars and phosphates. The most common nitrogenous bases found naturally in DNA are adenine (A), cytosine (C), guanine (G), and thymine (T). In naturally occurring RNA, T is replaced by uracil (U). Bases in DNA and RNA occur in ordered sequences that have the informational content required for the expression of genes, and in the case of RNA, for additional structural and metabolic functions. The ordered sequence of nucleotides is known as the *primary* structure, defined as an ordered sequence S = ( $b_1, b_i, ..., b_n,$ ) of length *n* with bases $b \in$ {A, C, G, U}. The sequence S is the primary input needed for *secondary* structure prediction.

Qknot is the term we use to denote the combination of variable (x) enumeration of stacked base pairs and Q matrix generation followed by optimization of xQx. Given an input sequence S of length n, Qknot consists of the steps outlined in the pseudocode below. The minimum distance parameter `min_d` refers to the minimum number of unpaired bases allowed between a closing base pair and is conventionally set to four. The `energy_weights` refer to the stacked quartet energies from (Turner & Mathews, 2009). In our analysis, we utilized the commercial solver AlphaQUBO (Meta-Analytics, 2020) as the `QUBO_search` method for returning a set of solutions T.

```
P ← Determine_feasible_pairs ( S, min_d )
R ← Generate_set_of_stacked_quartet_variables ( P, energy_weights )
Q ← Generate_row_col_value_file ( R, M⁺, M⁻, Mᴾ )
T ← QUBO_search ( Q, time_limit )
Recommended_Solutions ← Post_Process ( T )
```

**AlphaQUBO.** AlphaQUBO is a commercial solver from (Meta-Analytics, 2020) that incorporates metaheuristic research in Tabu Search (Glover, 1997), Path-Relinking and Scatter Search (Glover, 1997) as well as other classical methods for solving QUBO problems. There is no custom problem knowledge incorporated into the solution methodology, hence it is a generic solver, albeit a powerful one. The current version of AlphaQUBO solves any density QUBO problems with hundreds of thousands of variables and is available as a cloud-based solution. The largest problems solved in this research were for RNA of length 2904 generating a QUBO with 104K variables. AlphaQUBO uses the specified number of available processors and has multiple parameters affecting the speed and accuracy of the search process.

The solution process implemented in this research generated six AlphaQUBO threads in parallel, each one using two processors, hence at least 12 processors are required. A set of solutions T from each thread is stored in a solution file that is post processed after the time limit has been reached. Post processing consists of removing any base pairs that although they were feasible are not probable in terms of commonly encountered evolutionary pathways, e.g., an isolated stacked quartet as part of a sequence of unpaired bases. Additionally, the set of solutions T is analyzed to identify the most similar structure according to the novel metric Weighted Most Similar Structure described below.

**Post Processing and Data Analysis**. The objective function of the model presented relies on the thermodynamic energies of nested pairs described in Figure 3 which is used in many other secondary prediction programs [cite]. The basic premise is that the structure with the minimum free energy (MFE), or in our model the maximum bound energy, is the most stable and therefore represents the final folded state of the RNA. Testing indicates a high correlation between MFE and Qknot objective function values. However the structures determined by laboratory techniques such as X-ray crystallography and NMR indicate that the optimal most stable structure found in silico is not necessarily the structure the RNA forms in vivo and suboptimal solutions to MFE calculations may be closer to the actual structure (Zuker, et al., 1991). To address the issue of selecting only one, or very few, structures from the set T of discovered solutions to use for comparison to a benchmark, we introduce the metric Weighted Most Similar Structure (WMSS).

If a single solution is required, as for benchmark comparisons, then the set of solutions T is analyzed and a single best solution chosen according to WMSS. We analyze T by comparing all elements $t_i \in T$ using the similarity measures Matthews Correlation Coefficient (MCC) calculated from the number of True Positives (TP), True Negatives (TN), False Positives (FP) and False Negatives (FN) as compared to a benchmark. MCC is like F-score (or $F_1$) which varies from 0 to 1, while MCC varies from -1 (complete disagreement) to 0 (random agreement) to 1 (complete agreement). MCC is calculated as:

$$\text{MCC} = \frac{TP \times TN - FP \times FN}{\sqrt{(TP+FP)(TP+FN)(TN+FP)(TN+FN)}}$$

In order to calculate WMSS, each solution $t_i \in T$ is compared to every other solution in T to create an intra-solution similarity matrix from which an average and median MCC is generated for each $t_i \in T$.

```
for i = 1 to number of solutions in T
    set ti as the comparison benchmark solution
    for j = 1 to number of solutions in T, j <> i
        Similarity_matrix[i, j] ← calculate MCC(si, sj)
```

The average of all elements in each column i in the Similarity_matrix corresponds to the average MCC where $t_i$ is the solution used for comparison. The highest average indicates the solution that has the most common structure. However, these averages may be close to one another and a larger objective value indicates better energy, hence a weighting based on a normalized objective value is used. The results for weighted average and weighted median based on the following formula for each solution *i* will be a number between -1 and 1.

```
Weighted_Average_MCCi = Normalized_Objective_Valuei * Average_Similarityi
Weighted_Median_MCCi  = Normalized_Objective_Valuei * Median_Similarityi
```

Finally, the two measures for each solution *i* are added together to produce the WMSS which is a number between -2 and 2. The solutions with the largest WMSS are selected as the best representative among the set of solutions found and are used for comparison to an NMR or X-ray crystallography benchmark. In addition, AlphaQUBO outputs the solution pathway so that the intermediate structures leading to the final one can be analyzed as a folding solution pathway.

**Solution Pathways**. The article (Castro, et al., 2020) on RNA folding landscape states that the focus on MFE draws attention away from structures whose energies are suboptimal, and this may hinder new findings for two reasons. The first is that the dynamic and crowded intracellular environment is likely to prevent RNA from folding into its MFE structure. Secondly, inaccuracies in the energy function used by the RNA secondary structure prediction method may produce a false MFE. For these reasons, it is vital to examine the larger set of folds possible in order to fully understand the structural diversity of a given sequence. In other words, analyzing the folding pathways that occur during the RNA folding process is crucial. We utilized the output from AlphaQUBO to generate the folding pathway for RFA_00429 from the RNA Strand Database, The results are summarized in Figure 5.

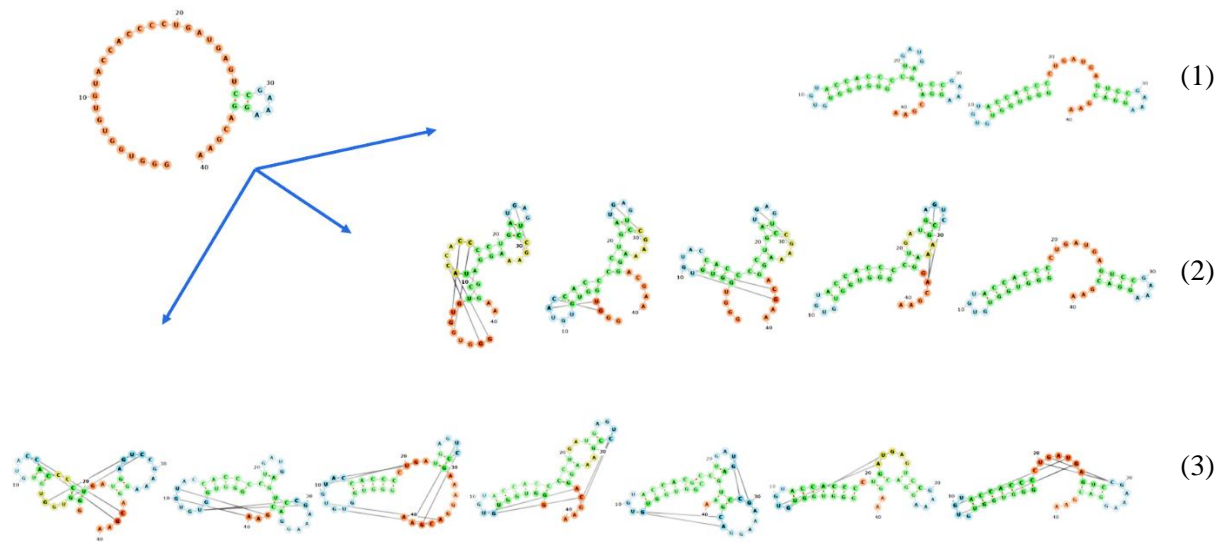

Figure 5. Progression of solutions towards most stable configuration (left to right) as reported for molecule RFA_00429 with 2D folding only (1), 3D pseudoknots penalized (2), and pseudoknots marginally penalized (3). The three paths all start with the same structure (upper left).

**Computational Experiments**. A set of pseudoknotted problems from RNA STRAND with no non-canonical base pairs were evaluated using AlphaQUBO as the solver. All problems were run on a Windows 10 PC with i9 3.7 GHz processor and 64 GB RAM.

Table 1 summarizes the results of the testing on a sample of the molecules from RNA STRAND database (RNA STRAND Database, 2008). All selected molecules are structurally verified by either NMR or X-ray crystallography with pseudoknots with length over 28 nt and no non-canonical pairs. There are 760 RNA molecules in the database meeting these criteria and we plan on testing them all in a future publication.

As the length increases the number of feasible folds increases as do the possibilities of optimal and near optimal energy states. Thus, MCC values tend to decrease as length increases. We used the structure with the highest WMSS in order to compute MCC. An MCC of 0.3 to 0.39 is considered a moderate correlation, over 0.4 to 0.69 is considered a strong correlation and above that very strong. Thus in this sample Qknot provides very good correlation with laboratory benchmarks.

As the length increases, the size of the QUBO instance increases approximately $O(n^2)$, however longer RNA molecules tend to pair locally, usually within 200 nt of each other.  This helps limit the combinatoric expansion.   For problems over 500 nt, the QUBO model limits the base pairs to be within 250 nt of each other.   Another factor affecting Q size are the number of feasible base pairs which is dependent on the organization of the bases in the linear sequence. Less than 10% of the RNA meeting our criteria have length > 1500 nt and most RNA have length less than 1000 nt.

| RNA name | length (nt) | # variables | time (sec) | MCC |
|---|---|---|---|---|
| PDB_00930 | 28 | 22 | 0.001 | 1 |
| PDB_00124 | 32 | 44 | 0.001 | 0.76 |
| RFA_00429 | 40 | 78 | 0.002 | 0.82 |
| PDB_00239 | 41 | 81 | 0.01 | 0.66 |
| PDB_01047 | 46 | 134 | 0.01 | 1 |
| PDB_01201 | 48 | 159 | 0.03 | 1 |
| RFA_00632 | 91 | 590 | 0.3 | 0.59 |
| ASE_00001 | 262 | 4083 | 23 | 0.49 |
| ASE_00005 | 341 | 6437 | 180 | 0.45 |
| PDB_00072 | 545 | 19900 | 1000 | 0.48 |
| PDB_00805 | 968 | 32934 | 554 | 0.51 |
| PDB_01261 | 968 | 27926 | 1746 | 0.64 |
| PDB_01101 | 1667 | 147407 | 900 | 0.39 |
| PDB_00029 | 2908 | 103980 | 1106 | 0.27 |

Table 1.   Testing results from sample of RNA STRAND database

**Conclusions**.   The 3-dimensional RNA folding prediction software Qknot was introduced along with results and analysis of testing from a sample of problems from the RNA STRAND database.   More detailed testing and results analysis will be available in another publication.   Qknot uses QUBO modeling in conjunction with soft penalties and rewards in the objective function to allow structures with pseudoknots to be predicted with very good correlation to benchmark structures. Qknot uses a state of the art QUBO solver that allows solving very large problems very quickly.   New similarity measures for working with a set of solutions, as opposed to a single solution, were introduced as a part of Qknot and show the approach is very promising as a flexible modeling framework for the difficult RNA folding prediction problem with pseudoknots.

**References.**